# Cell-Cycle-Associated Amplified Genomic-DNA Fragments (CAGFs) Might Be Involved in Chloroquine Action and Resistance in *Plasmodium falciparum*

Gao-De Li

Chinese Acupuncture Clinic, Liverpool, UK
Email: gaode_li@yahoo.co.uk

## Abstract

As a cheap and safe antimalarial agent, chloroquine (CQ) has been used in the battle against malaria for more than half century. However, the mechanism of CQ action and resistance in *Plasmodium falciparum* remains elusive. Based on further analysis of our published experimental results, we propose that the mechanism of CQ action and resistance might be closely linked with cell-cycle-associated amplified genomic-DNA fragments (CAGFs, singular form = CAGF) as CQ induces CAGF production in *P. falciparum*, which could affect multiple biological processes of the parasite, and thus might contribute to parasite death and CQ resistance. Recently, we found that CQ induced one of CAGFs, UB1-CAGF, might downregulate a probable *P. falciparum* cystine transporter (Pfct) gene expression, which could be used to understand the mechanism of CQ action and resistance in *P. falciparum*.

## Subject Areas

Cell Biology, Molecular Biology, Molecular Pharmacology

## 1. Introduction

Chloroquine (CQ) was developed in 1930's, and has been the mainstay of ma-laria chemotherapy for many decades. CQ resistance in *Plasmodium falciparum* first appeared in Thailand in 1957 and now is widespread in nearly all *P. falciparum* malaria endemic areas of the world [1]. Although CQ's value in malaria chemotherapy has diminished due to CQ resistance in *P. falciparum*, CQ's story continues as in recent years CQ has been found to enhance efficacy of various anticancer drugs [2], signifying renaissance of this old miracle drug.

However, to date, the mechanism of CQ action and resistance in *P. falciparum* remains controversial. Briefly, there are two main hypotheses regarding the mechanism of CQ action and resistance in *P. falciparum*, one could be named as the food-vacuole hypothesis that states that CQ's action site is the food vacuole in which toxic heme resulted from CQ's blockade of hemozoin formation kills malaria parasite and the food-vacuole membrane proteins, pfmdr1 and pfcrt, are responsible for CQ resistance [3]; another could be named as the nucleus hypothesis that claims that CQ's action site is the nucleus in which CQ causes DNA damage through DNA intercalation or apoptosis [4] [5] and defective DNA repair and alteration in DNA damage regulation might also contribute to CQ resistance [6] [7]. We once proposed that the key site of CQ action and resistance development is the nucleus of malaria parasite [8] and *P. falciparum* CQ resistance marker protein (Pfcrmp) might be the CQ's target in the nucleus [9] [10]. Recently, we reported that *P. falciparum* produces cell-cycle-associated amplified genomic-DNA fragments (CAGFs) at certain points of its intraerythrocytic cycle, and CQ can induce CAGF production [11]. In this paper, we present the results of further analysis of these findings, which might contribute to understanding the mechanism of CQ action and resistance in *P. falciparum*.

## 2. The Hypothesis

Based on further analysis of our published experimental results, we propose that the mechanism of CQ action and resistance in *P. falciparum* might be closely linked with cell-cycle-associated amplified genomic-DNA fragments (CAGFs, singular form = CAGF) as CQ induces CAGF production in *P. falciparum*, which could affect multiple biological processes of the parasite, and thus might contribute to parasite death and CQ resistance.

## 3. Background of Discovery of Cell-Cycle-Associated Amplified Genomic-DNA Fragments (CAGFs)

Cell-cycle-associated amplified genomic-DNA fragments (CAGFs) were discovered purely by chance [11]. Briefly, about 18 years ago, in order to check whether CQ could cause DNA damage in *P. falciparum*, both CQ-sensitive HB3 isolate, and CQ-resistant K1 isolate were treated with different doses of CQ. Parasite samples were harvested at 2 h, 6 h, 12 h, and 24 h intervals after the start of CQ treatment and then genomic DNA from the harvested samples was isolated. Using the isolated genomic DNA, we performed arbitrarily-primed PCR with one single arbitrary primer, named as AP1. Initially, we thought that if there was no DNA damage, the PCR banding patterns should be exactly the same in all samples, and if there was DNA damage the number of PCR bands would be reduced as DNA damage means that quantity and quality of DNA templates are reduced. As for the control groups, undoubtedly, the PCR banding pattern should remain the same in all samples. However, contrary to our anticipation, the PCR banding patterns in both control and CQ treated groups exhibited cell-cycle dependent changes, which was caused by two unstable bands, UB1 and UB2. In the control groups of both HB3 and K1 isolates, UB2 and UB1 appeared at 6 h and 12 h, respectively, whereas in the CQ-treated groups, both UB1 and UB2 appeared in the 2h samples of both isolates. Furthermore, both UB1 and UB2 in the CQ-treated groups were much clearer than those in the control groups, indicating that CQ induces both UB1 and UB2 production in both isolates. Appearance and disappearance of UB1 and UB2 at different points of the parasite's intraerythrocytic cycle suggest that the copy number of UB1- and UB2-related DNA templates increased and decreased accordingly during cell-cycle progression. This is why we named UB1- and UB2-related templates as UB1- and UB2-related CAGFs which can be further simplified as UB1-CAGF and UB2-CAGF. Here, the word, "amplified", refers to multiple copies, and does not mean gene or DNA amplification because if UB1- and UB2-related templates increased due to DNA amplification, the amplified templates would be incorporated into the genome and would not disappear few hours later. Besides, DNA amplification can not be induced 2 h after the start of CQ treatment. Therefore, we speculate that CAGFs are single-stranded DNA (ssDNA) fragments or molecules that are released from the genome into the nucleoplasm.

Mechanism by which CAGFs are produced and released from the genome is unknown, but we have proposed three possible mechanisms which include DNA-polymerase dependent DNA to DNA transcription [11] [12], endonuclease dependent transcript cutout [13], and genoautotomy (genome "self-injury") [14]. Originally, we speculated that both endonuclease dependent transcript cutout and genoautotomy mainly cut DNA at non-coding DNA region, but now we think that active transcription region might be the main target for both of them because recently we found UB1 sequence belongs to the coding DNA sequence of a gene. In the following section, the UB1-CAGF related gene's function and its association with CQ action and resistance in *P. falciparum* will be described in more detail.

If CAGF mainly comes from the coding DNA region of a gene, there will be enormous implications for biological science. First, since CAGF originates from coding DNA region, its sequence should be complimentary to either sense or antisence-strand sequence of the transcription region, the gene located within this transcription region could be named as CAGF-related gene. Thus, as a ssDNA molecule, CAGF can downregulate CAGF-related gene expression through either antisense or antigene or both strategies [15]; second, if CAGF generation is through endonuclease dependent transcript cutout [13] or genoautotomy (genome "self-injury") [14], a transient ssDNA gap caused by releasing CAGF will be formed at transcription region before complete repair of the gap. Both CAGF and ssDNA gap belong to ssDNA that is more prone to mutations than double-stranded DNA [16]. Furthermore, certain small fragments of CAGFs could serve as intrinsic mutation-contained primers that might be involved in natural site-directed mutagenesis [17]. Therefore, CAGF generation during cell cycle might contribute to drug resistance and evolution; third, if CAGFs can be released from cell as cell-free DNA (cfDNA) into blood stream in mammals, they might serve as mobile genetic elements [18] involved in horizontal or vertical genetic transfer. Since CAGFs might carry both genetic and epigenetic alterations in a gene, they might be involved in inheritance of acquired characteristics through vertical genetic transfer. Taken together, CAGFs might have a wide range of biological functions, but their main function is downregulation of CAGF-related gene expression, which is required for normal cell cycle progression.

## 4. Linkage between CAGF and CQ Action and Resistance in *P. falciparum*

As mentioned above, CAGF's main function is downregulation of CAGF-related gene expression. In our published paper, two CAGFs, *i.e.* UB1-CAGF and UB2-CAGF, were reported, which means that there are two CAGF-related genes, UB1-CAGF related gene and UB2-CAGF related gene, which could be downregulated by UB1-CAGF and UB2-CAGF, respectively. However, till now, only partial UB1-CAGF has been cloned and sequenced, therefore, in this section, we mainly focus on the function of a gene downregulated by UB1-CAGF, which might contribute to CQ action and resistance in *P. falciparum*.

About 18 years ago, we cloned and sequenced UB1 band, the obtained sequence shares 80% amino acid similarity with an unknown gene of *Escherichia coli* (P32667). The accession number of P32667 has now been changed to P0ABT8 in UniProtKB. Recently, we found that P0ABT8 has been named as probable cystine transporter YijE, which suggests that UB1-CAGF comes from the transcription region which might harbour a probable *P. falciparum* cystine transporter (Pfct) gene. This is a surprising new finding because the gene has not yet been reported in *P. falciparum* so far and might be involved in glutathione (GSH) synthesis.

The probable cystine transporter YijE gene encodes a membrane protein that is involved in response to cystine in *E. coli*, and overexpression of the protein confers tolerance to excess cystine [19]. In eukaryotic cells, the cystine transporter is also called cystine/glutamate antiporter that takes in extracellular cystine in exchange for intracellular glutamate [20] [21]. *P. falciparum* belongs to eukaryote and thus Pfct should function as a cystine/glutamate antiporter. Since uptake of cystine is required for intracellular GSH synthesis [22], Pfct might play a critical role not only in intracellular GSH synthesis, but also in GSH maintenance as cystine deprivation can induce GSH efflux and degradation [23].

Cystine used for GSH synthesis in *P. falciparum* might come from haemoglobin digestion in the food vacuole of the parasite. Therefore, it is most likely that Pfct is located in the food-vacuole membrane and transports cystine out of the food vacuole for GSH synthesis. The synthesized GSH will then be distributed in all intracellular compartments including the nucleus to exert its antioxidant and other biological effects.

CQ action and resistance in *P. falciparum* might be closely associated with Pfct due to two reasons: first, according to our published paper, UB1 and UB2 bands in the 12 h-HB3 control is much clearer than that in the 12 h-K1 control (Figure 1), suggesting that CQ-sensitive HB3 isolate produces more UB1- and UB2-CAGFs than CQ-resistant K1 isolate. Less UB1-CAGF induction in K1 isolate means more Pfct expression and more GSH synthesis in K1 isolate, which is consistent with previous report that levels of total GSH was higher in CQ-resistant *P. falciparum* Dd2 isolate than in CQ-sensitive *P. falciparum* 3D7 isolate [24]. Interestingly, increased GSH synthesis is also found in various drug-resistant cancer cell lines [25]; second, CQ at low dose (25 nM) can effectively induce both UB1- and UB2-CAGFs at early stage (2 h) of intraerythrocytic cycle of both HB3 and K1 isolates, but there is difference between two isolates in sustainability of UB1- and UB2-CAGFs. In K1 isolate, both UB1- and UB2-CAGFs disappeared at 12 h, while in HB3 isolate, only UB2-CAGF disappeared at 12 h, indicating more degradation of UB1-CAGF in K1 isolate than in HB3 isolate (Table 1).

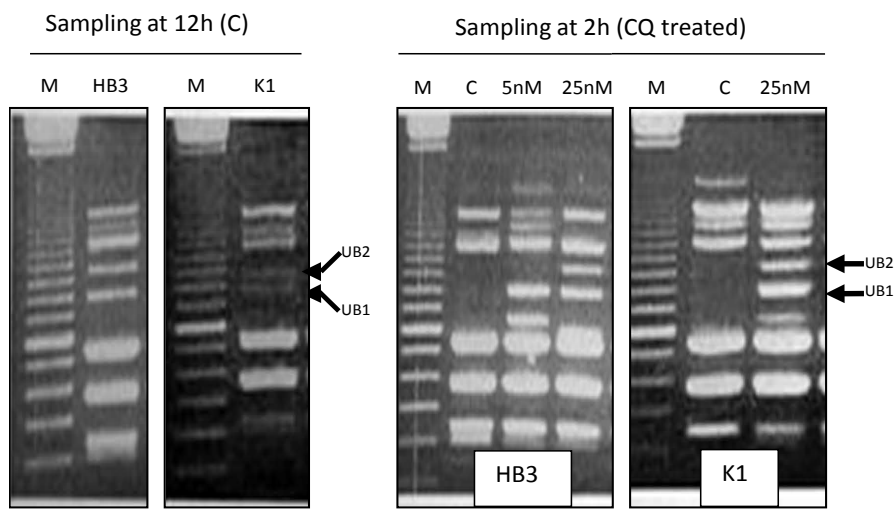

**Figure 1.** Arbitrarily-primed PCR banding patterns from CQ-sensitive HB3 and CQ-resistant K1 isolates (from Li, 2016 [11]). M = 100-bp-DNA ladder marker; C = control; CQ = chloroquine; UB1 size ≈ 1000 bp; UB2 size ≈ 1150 bp.

**Table 1.** Appearance (+) and disappearance (−) of UB1 and UB2 in the PCR products from the genomic-DNA samples of HB3 and K1 isolates using * indicating difference between the two isolates (from Li, 2016 [11]).

| Isolates | Treatment | PCR bands | 2 h | 6 h | 12 h | 24 h | Total + |
|---|---|---|---|---|---|---|---|
| HB3 | Control | UB1 | − | − | + | − | 4 |
|  |  | UB2 | − | + | + | +* |  |
|  | CQ (25 nM) | UB1 | + | + | +* | + | 6 |
|  |  | UB2 | + | +* | − | −* |  |
| K1 | Control | UB1 | − | − | + | − | 3 |
|  |  | UB2 | − | + | + | −* |  |
|  | CQ (25 nM) | UB1 | + | + | −* | + | 5 |
|  |  | UB2 | + | −* | − | +* |  |

Table 1 shows that more + in HB3 isolate than in K1 isolate, which suggests that CQ induced UB1-CAGF downregulated Pfct gene expression in HB3 isolate more severely than in K1 isolate. Thus, it can be concluded that overall GSH levels in K1 isolate were higher than those in HB3 isolate during CQ treatment. Since GSH plays an important role in antioxidant defence, low GSH levels indicate low antioxidant capacity, which will make the parasite more vulnerable to the attack of reactive oxygen species (ROS) that are constantly produced during oxidative metabolism and serve as a major contributor to oxidative damage of carbohydrates, nucleic acids, lipids, and proteins [26]. Furthermore, GSH depletion can elicit apoptosis and necrosis [27] and depletion of nuclear GSH impairs cell proliferation [28]. Taken together, CQ might kill CQ-sensitive *P. falciparum* through inducing UB1-CAGF that downregulates Pfct expression, which decreases GSH synthesis, and thus reduces antioxidant capacity, making the parasite more vulnerable to ROS attack. Among the various oxidative damages caused by ROS, DNA damage might contribute largely to parasite death. Less production and more degradation or more extracellular release of UB1-CAGF might contribute to CQ resistance in *P. falciparum*.

Theoretically, if downregulation of Pfct expression by UB1-CAGF is closely linked with CQ action, the timing of CQ induced UB1-CAGF should be consistent with the timing of parasite stages on which CQ acts. CQ induced UB1-CAGF appeared as early as 2 hours after the start of CQ treatment, which means that the parasite development was at the early ring-form stage [29]. If CQ induces UB1-CAGF in the early ring-form stage parasite, Pfct expression in this stage will be greatly downregulated, which means less GSH synthesis, and reduced antioxidant capacity, and therefore making the stage parasite more vulnerable to ROS attack. But previous research shows that the late stages, *i.e.* trophozoite and schizont, are considerably more sensitive to CQ than ring-stage parasites, which is coincident with elevated glucose consumption and nucleic acid synthesis [30]. One explanation for this contradiction is that the ring-form stage parasite might produce less ROS than late stage parasite due to less glucose metabolism as ROS is involved in regulation of glucose metabolism and glucose can induce ROS production [31] [32]. The main difference between HB3 and K1 isolates in CQ's induction of UB1-CAGF is that CQ induced UB1-CAGF was sustained in 12 h-sample of HB3 isolate, but disappeared in 12 h-sample of K1 isolate (Table 1). Since 12 h-sample is equivalent to late ring-form stage sample, it can be speculated that GSH levels in the late ring-form stage of K1 isolate is higher than those in the late ring-form stage of HB3 isolate, which can be used to understand the reason why K1 isolate is insensitive to CQ treatment. A clinical research has shown that CQ acts principally on the large ring-form and mature trophozoite stages of *P. falciparum* [33], which is consistent with our speculation. In conclusion, downregulation of Pfct expression by UB1-CAGF is closely linked with CQ action and resistance in *P. falciparum*.

Till now, we don't know what gene is downregulated by UB2-CAGF, however, since CQ can also induce UB2-CAGF production, it is possible that UB2-CAGF related gene's function might also contribute to CQ action and resistance in *P. falciparum*. Besides, in addition to induction of UB1- and UB2-CAGFs, CQ might induce other CAGFs in *P. falciparum*, which needs further investigation.

We once proposed that *P. falciparum* CQ resistance marker protein (Pfcrmp) was a CQ's target in the nucleus [9] as the genetic alterations in Pfcrmp gene were correlated well with CQ resistance phenotype after checking more than 100 isolates from three different regions [10]. Pfcrmp's function is unknown, but might be involved in enhancement of histone H3-modulated chromatin structure which is required for diverse bioprocesses, such as transcription, DNA repair, chromatin compaction during cell division and apoptosis [34]. The reason why this assumption is made is because Pfcrmp contains a DNMT1-RFD domain that is a potent histone H3 binding domain [35]. Since histone H3 phosphorylation is involved in ROS-mediated DNA damage and cell death [36] [37], it is possible that binding of CQ to wild-type Pfcrmp might aberrantly induce chromatin-structure changes through interaction with histone H3, which will make the chromatin DNA vulnerable to ROS attack. CQ might not induce the same chromatin-structure changes through interaction with mutant Pfcrmp, which contributes to CQ resistance.

## 5. Conclusions

Discoveries of CAGFs and Pfcrmp in *P. falciparum* strongly support the idea that the key site of CQ action and resistance development is within the nucleus of malaria parasite. CQ has diverse therapeutic actions, which might be closely linked with CQ's induction of various CAGFs. Since CAGFs might also be produced and induced by CQ in other eukaryotic cells, such as cancer cells, further investigation into this area may not only contribute to understanding the mechanism of CQ action and resistance in both *P. falciparum* and cancer cells, but also helps to design novel drugs and strategies to tackle drug resistance in both *P. falciaprum* and various cancer cells.

Downregulation of a probable Pfct gene expression by UB1-CAGF is a novel mechanism by which intracellular GSH levels are regulated. It has been proved by many experimental results that GSH levels are associated with CQ action and resistance in the rodent malaria species *P. berghei*, but the similar convincing evidence in *P. falciparum* has not yet been obtained [38]. No doubt, UB1-CAGF modulated Pfct expression supports the view that GSH plays an important role in CQ action and resistance in *P. falciparum*. Unfortunately, 18 years ago, we didn't submit the UB1 DNA sequence to GenBank and now the sequence has gone missing. Luckily, the two specific primers we published previously were designed based on the sequence [11], which can be used to easily clone this unreported Pfct gene.

CAGFs belong to extrachromosomal ssDNA molecules and their copy numbers are increased after CQ treatment [11], which distinguishes them from double-stranded extrachromosomal circular DNA (eccDNA) molecules found in all eukaryotes [39] [40]. Although eccDNA molecules can be generated by sublethal drug exposure [41] and levels of total eccDNA molecules have been elevated after exposure of organism to genotoxic agent [42], the copy number of every single eccDNA molecule can not be increased after exposure to genotoxic agent because the formation of eccDNA molecules is through excision of chromosomal sequences and no de-novo DNA synthesis is involved [43]. The eccDNA-related excision of chromosomal sequences could be considered as a kind of geno-autotomy in which nonlethal double-stranded genomic-DNA fragments are excised from chromosomes [14].

In conclusion, CQ's induction of CAGFs in *P. falciparum* is a very important discovery that may open up an entirely new field of research.

## Conflict of Interest

The author declares that there is no conflict of interest regarding the publication of this paper.

## List of Abbreviations Used in This Article

| | |
|---|---|
| CQ | chloroquine |
| H | hour |
| Pfct | *Plasmodium falciparum* cystine transporter |
| GSH | glutathione |
| ROS | reactive oxygen species |
| CAGF | cell-cycle-associated amplified genomic-DNA fragment |
| UB1 | unstable band (UB) 1 in PCR products |
| UB2 | unstable band (UB) 2 in PCR products |
| UB1-CAGF | UB1-related CAGF |
| UB2-CAGF | UB2-related CAGF |
| Pfcrmp | *Plasmodium falciparum* chloroquine resistance marker protein |
| DNMT1-RFD | DNA (cytosine-5)-methyltransferase 1, replication foci domain |
| eccDNA | extrachromosomal circular DNA |